\documentclass[11pt]{article}
\usepackage{pdproc}

\usepackage{graphicx}% Include figure files
\usepackage{amssymb}% Define \lesssim and \gtrsim
\usepackage{bm}% Bold math

\newcommand{\be}{\begin{eqnarray}}      \newcommand{\ee}{\end{eqnarray}}   
\newcommand{\ba}{\begin{array}}         \newcommand{\ea}{\end{array}} 
\newcommand{\ct}[1]{\cite{#1}}
\newcommand{\rf}[1]{~(\ref{#1})}        

\newcommand{\lb}[1]{\label{#1}}

\newcommand{\lf}{\left}     \newcommand{\rt}{\right}
\newcommand{\fr}{\frac}     

\newcommand{\dl}{\delta}    \def\d{\delta}
\newcommand{\pd}{\partial}  \newcommand{\D}{\Delta}

         \newcommand{\g}{\gamma}
 
\def\t{\tau}
\newcommand{\Nb}{\nabla}       
\newcommand{\Hc}{{\mathcal H}}
\newcommand{\Om}{\Omega}
\newcommand{\sg}{_{\gamma}}  
\newcommand{\snu}{_{\nu}}           
           
\newcommand{\gdec}{_{\gamma\,\rm d}}           
\newcommand{\eq}{_{\rm eq}}

\begin{document}

\title{\large Robust Signatures of the Relic Neutrinos in CMB\footnote[1]{
Based on the proceedings for the 10th 
Int.\ Symposium on Particles, Strings and Cosmology 
(PASCOS 04) at Northeastern University, Boston, August 16-22, 2004.}
}

\author{\normalsize Sergei Bashinsky}
\address{\it International Centre for Theoretical Physics\\ 
          Strada Costiera 11, 31014 Trieste, Italy\\
          \rm E-mail: bashinsky@ictp.trieste.it %\\
          %\rm(\today)
          }

\abstract{
When the perturbations forming the acoustic peaks of the cosmic 
microwave background (CMB) reentered the horizon
and interacted gravitationally with all the matter, neutrinos presumably 
comprised $41\%$ of the universe energy.  CMB experiments have
reached a capacity to probe this background of relic neutrinos.
I review the neutrino impact on the CMB at the onset of the acoustic 
oscillations.
The discussion addresses the underlying physics, 
robustness or degeneracy of the neutrino imprints with changes of 
free cosmological parameters, and non-minimal models for the 
unseen radiation sector with detectable signatures in CMB
anisotropy and polarization.
}

\renewcommand{\thefootnote}{\alph{footnote}}
\setcounter{footnote}{0}

\section{Introduction}
\label{sec:intro}
The counts of solar, atmospheric, reactor,
and accelerator-produced neutrinos provide a solid evidence of physics beyond 
the Standard Model of particle physics. 
They can be explained by neutrino flavor mixing by non-zero masses,
subject to tightening upper cosmological bounds\ct{m_nu_constr}. 
These upper mass bounds, however, rely on the assumption that the 
neutrino sector hides no additional surprises.  Specifically, that 
a neutrino per photon ratio has been fixed since the decoupling of the 
active neutrinos and the subsequent $e^+e^-$ annihilation.
Or that neutrinos do not couple to unseen 
fields\ct{mnu_decay_bcm,mnu_decay_chk}.
These assumptions are verifiable by cosmological observations. 

Perturbations in the cosmic microwave background (CMB) carry
imprints of any species which contributed to 
the universe density when the perturbations reentered the horizon\footnote{
  We say that a perturbation mode ``reentered the horizon'' 
  when the Hubble scale $H^{-1}$ exceeded the mode wavelength.
}.
Some of these imprints cannot be mimicked by changes of the
standard cosmological parameters.
Moreover, they reveal the {\it internal interactions\/} 
of species which couple only gravitationally to the visible sector.
The big bang nucleosynthesis (BBN) yield of light elements, 
notably He$^4$, is as well sensitive to all the dominant species
in the radiation epoch, but at high redshifts $z\sim 10^{9}-10^{8}$
($\sim 100$\,keV).
The CMB peaks, on the other hand, probe the late radiation era\footnote{
  The redshifts probed by the CMB multipoles with $l\le 1500$ [$\lesssim3000$] 
  are $z_{l,\rm entry}\lesssim 3\times 10^4$ [$\lesssim 6\times 10^4$], 
  assuming the concordance $\Lambda$CDM model.
}
$z\sim 10^4-10^3$ ($\sim 10-1$\,eV).

The first WMAP data release\ct{WMAPBennett} 
places the effective number of relic 
neutrinos in a $\mbox{2-$\sigma$}$ range\ct{N_nu_CMB} 
$1 \lesssim N\snu \lesssim 7$,
mildly improved\ct{N_nu_CMB} by complementary probes. 
The range will be narrowed by an order of magnitude and 
more as the CMB sky is mapped at smaller angles:
at 1\,$\sigma$, to\ct{Nnu_projections,BS} 
$\D N\snu\approx 0.2$ by PLANCK
and up to $\D N\snu\approx 0.05$  
by more advanced proposed missions.

\section{Imprints of neutrino gravity on CMB}
\lb{sec_imprints}

The CMB fluctuations could not escape the gravity of relic neutrinos
during the radiation era, when neutrinos were among the dominant species.  
The neutrino fraction $R\snu\equiv {\rho\snu/\rho_r}$
of the total radiation energy density $\rho_r=\rho\sg+\rho\snu=\rho\sg/(1-R\snu)$
is $41\%$ in the standard scenario.
With the present cosmological bounds\ct{m_nu_constr} 
$m\snu\lesssim 0.14$ to $0.6$~eV, 
the standard neutrinos were all highly {\it relativistic\/} 
in the radiation era, prior to matter-radiation density equality at 
%\be
%z\eq+1 %\equiv \fr1{a\eq} 
% = (\rho_{c+b,0}/\rho_{\g,0})(1-R\snu).
%\lb{zeq}
%\ee
\be
\qquad
z\eq+1 %\equiv \fr1{a\eq} 
    = \fr{\rho_{c+b,0}}{\rho_{\g,0}}(1-R\snu)
       \simeq 3.3\times 10^3\, {\Om_{c+b}h^2 \over 0.14}
                        \,{1.69\over 1+0.23\,N\snu}.
\lb{zeq}
\ee
[Here, $c{+}b$ labels cold dark matter (CDM) plus baryons,
``$0$'' denotes the present values, and the last two fractions 
are close to unity in the standard $\Lambda$CDM + 3$\nu$ scenario.]
%$z\eq\approx 3.2\times10^3$ in the standard scenario.]
These neutrinos become non-relativistic below the redshift
$z_{m\snu}\!\approx m\snu/(3kT_{\nu,0})\simeq 200\,m\snu/(0.1\,{\rm eV})$,
{\it i.e.}, only after the CMB decoupling ($z\gdec\approx 1090$).

The growth of cosmological structure
in the matter era is also affected by neutrino mass density.
It carries signatures of a neutrino free streaming scale\ct{BondSzalay}, 
which by then is sensitive to $m\snu$. 
Although the low neutrino masses influence the perturbation dynamics only 
after the CMB last scattering, they may in future be probed by CMB
due to their impact on large-scale structure and induced by it 
CMB lensing\ct{Kaplinghat}.

The impact of relic neutrinos cannot be ignored in the 
search for primordial gravity waves (hence, probing the energy scale 
of inflation) through $B$-polarization of CMB\ct{BfromGravity}.  
Interaction of neutrino perturbations with tensor 
gravity waves suppresses the tensor amplitude on 
the scales entering the horizon in the radiation era ($l>200$)
and reduces the $B$-mode spectrum by as much as\ct{WeinbTensor} $36\%$.
Even on the largest angular scales neutrinos damp the tensor 
power by about\ct{WeinbTensor} $10\%$.

In the rest of this section we review the signatures of the abundant in the radiation 
era neutrinos in the CMB temperature anisotropy and E-polarization.

\subsection{Background effects}

According to Friedmann equation, higher neutrino density speeds 
up the Hubble expansion in the radiation era. 
For {\it fixed\/} density $\Om_{c+b}h^2$ ($\propto \rho_{c+b,0}$), 
the faster expansion would reduce the size of the CMB acoustic 
horizon.
In addition, a larger fraction of radiation density prior to the CMB 
last scattering would enhance the first acoustic peak
by a stronger early integrated Sachs-Wolfe 
effect from the increased proximity of the radiation-matter 
transition\rf{zeq} to the last scattering and by lesser suppression of 
the peak by matter perturbations\ct{Bash}\footnote{
  \lb{foot_driving}
  The accepted today
  interpretation of an order of magnitude larger
  CMB power on the scales entering the horizon before versus
  after the equality as the resonance driving of 
  radiation perturbations by their self-gravity is incorrect. 
  This can be demonstrated by the following example:  
  Suppose that long after the superhorizon freezout of perturbations 
  but before their horizon reentry in the radiation era, photons and neutrinos 
  had become subdominant to an {\it unperturbed\/} decoupled 
  component~$X$ with 
  the radiation equation of state $p_X=\rho_X/3$.
  In this scenario, in which the metric is unperturbed and no gravitational
  driving of CMB is expected, the acoustic CMB oscillations 
  after the horizon reentry would have {\it precisely\/} 
  the same amplitude as in the model dominated by a photon fluid.
  To the contrary, a similar construction for the matter era\ct{Bash} 
  shows that then 
  the freely collapsing CDM perturbations, generating non-decaying 
  gravitational potential, cause {\it 25-fold suppression\/} 
  of the CMB power~$C_l$ in the Sachs-Wolfe\ct{SachsWolfe} 
  matter era result.
}.

Although the change of the acoustic horizon
and the consequences of the shifted $z\eq$ are frequently 
mentioned in the literature, {\it neither\/} of these effects
can serve as a neutrino probe.
Indeed, our knowledge of other densities, in particular CDM, is likewise derived
from the observed correlations of cosmological inhomogeneities.  
Even when $\rho\snu$ is not equal to its assumed value, the angular  acoustic horizon scale
and~$z\eq$ remain essentially unchanged when at any~$z$
the ratio $\rho_b/\rho\sg$\footnote{ 
  The photon density $\rho\sg$ is fixed by COBE-measured\ct{T_COBE}~$T_{\rm CMB}$.
}, hence the photo-baryon sound 
speed, has the standard value
while the densities $\rho_{c+b}$, $\rho_{\rm dark}$ 
[and $\rho_{K,0}\equiv 3(1-\Om)H_0^2/(8\pi G)$ 
for non-flat models] are rescaled in proportion 
to~$\rho_r=\rho\sg+\rho\snu$.

The observed matter power $P_m(k/h)$ remains almost unchanged as well.\footnote{
  The popular rule of thumb ``LSS constrains $\Om_m h$'' assumes a fixed
  neutrino fraction of radiation density and fails in our situation. 
  Indeed, the physical quantity probed best by the 
  large scale structure is the ratio of the comoving Hubble 
  scale at the matter-radiation equality~$\Hc\eq^{-1}=(H\eq a\eq)^{-1}$
  to the present Hubble scale~$H_0^{-1}$. 
  With eq.\rf{zeq},
  $\Hc\eq/H_0\sim \Om_{c+b}^{1/2}(z\eq+1)^{1/2}\propto \Om_{c+b}h(1-R\snu)^{1/2}$.
  It is unchanged by the discussed (conformal) rescaling.
}
(Small changes are induced by variations in
$\rho_b/\rho_c$, in, so far undetected, signatures of neutrino masses,
and in the impact of neutrino perturbations, Sec.~2.2.)
There is a simple explanation: In the compared scenarios most of 
the characteristic time and distance intervals differ by the 
same factor.  
This conformal rescaling preserves measured angles and redshifts\ct{BS}.

Several consequences of the common density rescaling can still be 
observed. First, a different value of the Hubble 
constant $H^2_0\propto \sum_{a,K}\rho_{a,0}$.  
Constraints on~$H_0$, however, tend to be weak. 

Second, the scale of Silk damping\ct{Silk_damp} 
$\lambda_{\rm Silk}$ is determined by the mean time of photon 
collisionless flight~$\t_c=(an_e\sigma_{\g e})^{-1}$,
not affected by the density rescaling. 
(We use comoving scales and conformal times.)
A larger ratio of the damping scale to the Hubble radius
$\lambda_{\rm Silk}/\Hc^{-1}\sim \sqrt{\t_c\Hc}$
in models with greater~$\rho\sg$ leads to increased
damping of small-scale CMB anisotropies and helps  
to constrain~$N\snu$ (Table~III in Ref.~[7]).  %\ct{BS}).
However, the density of free electrons~$n_e$ around the photon 
decoupling, when only hydrogen was ionized, depends 
(for known~$\rho_b$) on the primordial helium 
abundance~$Y_p$: $n_e\equiv x_e n_H = x_e(1-Y_p)\rho_{b}/m_H$.
Taking for the present uncertainty $Y_p= 0.24\pm 0.01$\ct{Y_p}, 
$Y_p$ can be freely adjusted 
to rescale~$\lambda_{\rm Silk}$ in proportion to~$\Hc^{-1}$ 
for~$\D N\snu\lesssim 0.4$.
A detailed analysis\ct{BS} shows that the rescaling 
$1-Y_p\propto \rho_r^{1/2}$ yields almost degenerate~$x_e(z)$, 
hence equivalent CMB decoupling,
and unchanged $\lambda_{\rm Silk}/\Hc^{-1}$.

\subsection{Perturbation effects}
\lb{sec_nu_pert}

Finally, the conformal degeneracy of CMB and matter spectra in a faster 
expanding background is broken by the gravitational 
impact of neutrino perturbations.
Relative perturbation of the number density $n\sg$ of photons 
(coupled to baryons) evolves in the perturbed metric as
\be
\textstyle
\ddot d\sg + \fr{R_b \Hc}{1+R_b}\,\dot d\sg - \fr{1}{3(1+R_b)}\Nb^2d\sg 
          = \Nb^2(\Phi+\fr1{1+R_b}\,\Psi)\,.
\lb{dot_g}
\ee
Here $\Phi$ and $\Psi$ parameterize scalar metric perturbations  
$ds^2=a^2[-(1+2\Phi)d\tau^2+(1-2\Psi)dx^2]$, overdots denote the derivatives with 
respect to the conformal time~$\tau$, $\Hc\equiv \dot a/a$, 
$R_b=3\rho_b/(4\rho\sg)$, 
and 
\be
d\sg\equiv \fr{\dl \rho\sg}{\rho\sg+p\sg}-3\Psi
      = 3\lf(\fr{\d T\sg}{T\sg}-\Psi\rt)
\lb{dg}
\ee
(in the Newtonian gauge) is a general-relativistic generalization\footnote{
  $d\sg$ of eq.\rf{dg} is a {\it unique\/}\ct{Bash} perturbation variable which\, 
  {\bf a)} Reduces to $\d n\sg/n\sg$ on subhorizon scales\, 
  {\bf b)}~Freezes beyond the Hubble horizon and\, 
  {\bf c)} Experiences no gravitational driving on any scale in 
  a spatially uniform metric. (The unperturbed Robertson-Walker 
  metric can be imposed consistently by artificially populating 
  the universe with decoupled species~$X$ which density is negligible
  during the generation and superhorizon freezout of the studied perturbations, 
  and requiring that X subsequently become homogeneous in space and dominate
  the total energy density.  See footnote~$c$ and Ref.\,[12].)

  The corresponding gauge-independent expression for arbitrary 
species~$a$ reads: 
\be
 d_a\equiv 3\zeta_a 
    \equiv \fr{\dl \rho_a}{\rho_a+p_a}+3D+\nabla^2\epsilon,
\lb{d_a_ginv}
\ee
where $D$\ and $\epsilon $ parameterize scalar perturbations of
the spatial metric $\delta g_{ij}=2a^{2}[D\,\delta
_{ij}-(\nabla _{i}\nabla _{\!j}-\frac{1}{3}\delta _{ij}\nabla^{2})\,\epsilon]$. Various interpretations of this variable are:
in the spatially flat gauge ($\delta g_{ij}\equiv 0$)
it gives $\d n_a/n_a \equiv \delta \rho _{a}/(\rho _{a}+p_{a})$;  
in any gauge without shear ($\epsilon=0$) it describes
$\d n_{a,\rm coo}/n_{a,\rm coo}$, where $n_{a,\rm coo}$ 
is the particle number density with respect to coordinate 
rather than proper volume\ct{BS};
an equivalent variable\ct{WandsMalik} $\zeta_a=d_a/3$, a curvature perturbation
on the hypersurfaces of constant~$\rho_a$,
is in widespread use to describe superhorizon 
dynamics of several fluids.

   Eq.\rf{dot_g}\ct{BS} for CMB perturbations and similar 
   equations for other cosmological species\ct{BS,Bash} 
   have a series of advantages 
   over the traditional approaches to the CMB
   whenever horizon-scale gravity plays a role: 
   {\bf 1)}. The highest time derivatives
   of the dynamical perturbation variables now do not enter 
   the equations implicitly through time derivatives of the ``driving'' 
   potentials on the right hand side.
   This allows straightforward numerical integration
   of the equations in the 
   convenient and fixed Newtonian gauge. 
   {\bf 2)}. A direct connection\ct{Bash} to
   phase-space distributions in the perturbed metric.
   (See Refs.\,[8,19]   %\ct{BondSzalay,MaBert} 
   for the subtleties with the traditional variables.)
   {\bf 3)}. As stated above, a {\it change\/} of the perturbation measure $d_a$ 
   ($\zeta_a$) on any scale is caused {\it by a physical causal mechanism only\/}.
   This is not so for $\d T\sg/T\sg$, 
   $\d\rho_a/\rho_a$, and most of the other conventional variables, 
   unless one selects a gauge where they coincide or 
   proportional to~$d_a$ ($\zeta_a$).
   {\bf 4)}. Analytical solutions become simple\ct{BS} in
   this approach.\ct{Dod_book}  This facilitates analytical study of 
   the physical content of CMB features and leads to new results,
   including those described in this review. 
} 
of~$\d n\sg/n\sg$.

For the radiation era, when the CMB peaks enter the horizon, eq.\rf{dot_g}
is easily solved\ct{BS} in {\it real space\/} with the Green's function approach\ct{BB}:
\be
d\sg^{(\rm rad\, era)}(x,\t)=\fr{A\sg}{2}\d_{\rm Dirac}\!\lf(|x|-{\t\over \sqrt3}\rt)
                           +{\Phi+\Psi \over (x/\t)^2-1/3},
\lb{dg_real}
\ee
where we assume adiabatic initial conditions.
After Fourier transformation to $k$-space, the delta-function 
term describes the famous acoustic oscillations 
$A\sg \cos(k\t/\sqrt3)$ in the photon fluid with the speed of
sound $c_s=1/\sqrt3$.  Whenever the metric is perturbed
at the acoustic horizon, {\it i.e.\/} $\lf(\Phi+\Psi\rt)_{|x|=\t/\sqrt3}\not=0$,
the  small-scale singularity of the second term in eq.\rf{dg_real} 
contributes a sine component to the subhorizon oscillations in
Fourier space or, equivalently, {\it shifts\/} the oscillation {\it phase\/}.

As proven in Appendix~B of~[7], %\ct{BS}) 
the metric in the adiabatic Green's 
function\rf{dg_real} is perturbed beyond the acoustic horizon if only 
some perturbations physically propagate faster than the acoustic 
speed~$c_s$.  Among the standard cosmological species, only 
free-streaming relativistic neutrinos support a faster speed,
the speed of light.  
(Perturbations in early quintessence do as well.
They induce an even larger phase shift per equal density\ct{Bash}.)
This phase shift results in a non-degenerate additive
shift of all the acoustic peaks in~$C_l$, for either 
temperature or polarization spectra. 
For $N\snu\approx 3$, the positions of all 
the peaks change as $\pd l_{\rm peak}/\pd N\snu\approx -4$. 

In addition, neutrino perturbations somewhat suppress the magnitude of
the acoustic oscillations while enhance the matter power\ct{Hu_small,BS}.
With unknown primordial power, detection
of this effect requires precise measurement of the matter spectrum
and is difficult. Quintessence\ct{Quint} perturbations cause
an opposite change of the CMB to matter power ratio\ct{Q_LSS_suppression,Bash}. 
If CMB finds a non-standard~$N\snu$, this may, in principle, discriminate 
between contributions of sterile particles and early quintessence.

\section{Probing particle physics}

An apparent contribution to $N\snu$ may be provided by sterile neutrinos
(light neutral fields which mix with the active neutrinos).
Motivations to consider such fields and their status in the light
of the recent oscillation data and other constraints were reviewed in 
detail in Cirelli's contribution\ct{Cirelli}. 

The radiation density might also be enhanced by yet unknown light particles 
which neither couple nor mix to the visible sector and fell off the thermal
bath early, say, above the SUSY breaking scale.  At that high temperature
the radiation entropy was shared by a large number of relativistic species
and the decoupled particles would receive only a small fraction of the 
(conserved) total entropy.
Then the late-time contribution of the decoupled particles to the radiation 
density  relative to the contribution  of photons and active neutrinos
may 
be small 
and pass the current BBN constraints.

Since the phase shift of the CMB peaks
is {\it absent\/} when neutrinos do not free-stream,
CMB will probe a growing class of models where neutrinos 
couple or recouple to other fields.
The scenarios 
where recoupling takes place after BBN 
are not constrained\ct{Chacko03,mnu_decay_chk} 
by the BBN limits on $N\snu$.
For example, recoupling is expected if the small neutrino masses 
are generated by a coupling $l\nu_Rh\,\phi/M$ 
($l$ and $h$ are the standard lepton and Higgs doublets and $\phi$ is a new 
field) after low-energy spontaneous 
symmetry breaking $|\phi|+\d\phi$\ct{Chacko03}.  
Other neutrino mass models with neutrino coupling or recoupling have 
been considered\ct{mnu_decay_bcm,mnu_decay_chk,nu_composite}.
If coupled to a light field $\d\phi$, neutrinos generally 
annihilate\ct{mnu_decay_bcm,mnu_decay_chk} after they become 
non-relativistic.  Then neutrino mass is not 
bounded by the standard cosmological limits\ct{m_nu_constr}.
A changed phase of the acoustic peaks is a clean
signature\footnote{
  Neutrino coupling to a light 
  field~$\phi$ cannot be falsified, 
  as claimed by the authors of Ref.~[2],
  by the lesser large-scale structure growth due to
  the radiation-matter equality delay
  (caused by $\phi$ contribution to~$\rho_r$).
  At least, not until we find a non-cosmological probe of CDM density.
  See Sec.~2.1.
}  
of neutrino coupling or recoupling\ct{Chacko03}.

\section{Summary}

Today's upper cosmological bounds on neutrino masses\ct{m_nu_constr}
imply that neutrinos of all the three standard 
generations were relativistic in the radiation era.
Their energy density was comparable to that of photons,
and the neutrino gravity had major impact on the cosmological dynamics.
However, the induced changes of such important cosmological characteristics 
as the size of the acoustic CMB horizon, the redshift of
radiation-matter equality, and the Hubble radius at the onset 
of the structure growth have no observable 
consequences.  The corresponding angular scales and 
redshifts are unchanged in the scenarios where the
densities of the visible matter plus the dark matter and of the dark energy
are larger in proportion to the radiation density enhancement.  

This conformal degeneracy is lifted by Silk damping  
if the primordial helium abundance~$Y_p$ is fixed by 
independent spectrophotometric measurements.  
However, with the uncertainty 
$\D Y_p\sim 0.01$, an allowed adjustment of $Y_p$ results 
in degenerate Silk damping and degenerate CMB decoupling among the compared
scenarios for $\D N\snu <0.4$.

The gravitational impact of {\it neutrino perturbations\/} on the CMB 
temperature and polarization angular spectra is not degenerate with 
any of the standard parameters.
Free streaming of neutrino perturbations faster
than the acoustic speed causes a measurable shift 
of the acoustic oscillation phase.  The multipoles of 
all the acoustic peaks shift by an approximately same interval 
$\D l \approx - 4\,\D N\snu$ (at $N\snu\approx 3$).  This shift
does not occur or is reduced in the models where all or some 
of the neutrino types either had not decoupled or recoupled 
when the perturbations entered the horizon.  
(Similar changes might be induced by isocurvature modes,
which detection would be just as exciting.)

A canonical dynamical field (quintessence),
perturbations of which propagate at the speed of light, 
causes a similar uniform shift\footnote{
  For the quintessence
  energy density which in the radiation era equals the density of 
$N_Q$ effective fermions,
  $\D l \approx -11 \D N_Q$\ct{Bash}. 
}  
of the peaks. 
Early quintessence may be distinguished 
from extra light relativistic particles by accurate measurement of the 
CMB to matter power ratio.  Perturbations of a classical field and 
decoupled particles change this ratio in opposite directions.

Ultimately, neither dark matter nor dark 
energy entered the cosmological scene with strong theoretical motivation.  
The upcoming CMB data will testify whether or not the radiation 
epoch conceals its own surprises.

\section{Acknowledgments}
I thank the organizers of PASCOS 04, and I am grateful 
to S.~Shankaranarayanan and A.~Smirnov for their suggestions
during the preparation of this contribution.

\end{document}